\documentclass{stacs_proc}

\def\E{\mathbb{E}}
\def\F{\mathbb{F}}
\def\P{\mathbb{P}}
\def\T{\mathbb{T}}

\def\x{\textbf{x}}
\def\y{\textbf{y}}
\def\z{\textbf{z}}

\begin{document}

\title{Lower bounds for adaptive linearity tests}

%
%

\author{S. Lovett}{Shachar Lovett}
\address{Faculty of Mathematics and Computer Science
\newline The Weizmann Institute of Science, POB 26
\newline Rehovot 76100, Israel}
\email{Shachar.Lovett@weizmann.ac.il}


\thanks{Research supported by grant 1300/05 from the Israel
Science Foundation.}

\begin{abstract}
Linearity tests are randomized algorithms which have
oracle access to the truth table of some function f, and are
supposed to distinguish between linear functions and functions which
are far from linear. Linearity tests were first introduced by Blum,
Luby and Rubenfeld in [BLR93], and were later used in the PCP
theorem, among other applications. The quality of a linearity test is
described by its correctness c - the probability it accepts
linear functions, its soundness s - the probability it accepts
functions far from linear, and its query complexity q -
the number of queries it makes. 

Linearity tests were studied in order to decrease the soundness of
linearity tests, while keeping the query complexity small (for one
reason, to improve PCP constructions). Samorodnitsky and Trevisan
constructed in [ST00] the Complete Graph Test, and prove that no 
Hyper Graph Test can perform better than the Complete Graph Test. 
Later in [ST06] they prove, among other results, that no non-adaptive 
linearity test can perform better than the Complete Graph Test. 
Their proof uses the algebraic machinery of the Gowers Norm. 
A result by Ben-Sasson, Harsha and Raskhodnikova [BHR05] allows 
to generalize this lower bound also to adaptive linearity tests.

We also prove the same optimal lower bound for adaptive linearity
test, but our proof technique is arguably simpler and more direct
than the one used in [ST06]. We also study, like [ST06],
the behavior of linearity tests on quadratic functions. However,
instead of analyzing the Gowers Norm of certain functions, we
provide a more direct combinatorial proof, studying the behavior of
linearity tests on random quadratic functions. This proof technique
also lets us prove directly the lower bound also for adaptive
linearity tests.
\end{abstract}

\maketitle

\stacsheading{2008}{515-526}{Bordeaux}
\firstpageno{515}

\section{Introduction}\label{intro}
We study the relation between the number of queries and soundness of
adaptive linearity tests. A linearity test (over the field $\F_2$
for example) is a randomized algorithm which has oracle access to
the truth table of a function $f: \{0,1\}^n \to \{0,1\}$, and needs
to distinguish between the following two extreme cases:
\begin{enumerate}
  \item $f$ is linear
  \item $f$ is far from linear functions
\end{enumerate}

A function $f$ is called \emph{linear} if it can be written as
$f(x_1,...,x_n) = a_1 x_1 + ... + a_n x_n$, with $a_1,...,a_n \in
\F_2$. The agreement of two functions $f,g:\{0,1\}^n \to \{0,1\}$ is
defined as $d(f,g) = |\P_{\x}[f(\x)=g(\x)] - \P_{\x}[f(\x) \ne
g(\x)]|$. $f$ is far from linear functions if it has small agreement
with all linear functions (we make this definition precise in
Section~\ref{prelim}).

Linearity tests were first introduced by Blum, Luby and Rubenfeld in
\cite{BLR93}. They presented the following test (coined the BLR
test), which makes only 3 queries to $f$:
\begin{enumerate}
    \item Choose $\x,\y \in \{0,1\}^n$ at random
    \item Verify that $f(\x+\y) = f(\x) + f(\y)$.
\end{enumerate}

Bellare et al. \cite{BCH+96} gave a tight analysis of the BLR test.
It is obvious that the BLR test always accepts a linear function.
They have shown that if the test accepts a function $f$ with
probability $1/2 + \epsilon$, then $f$ has agreement at least $2
\epsilon$ with some linear function.

For a linearity test, we define that it has \emph{completeness} $c$
if it accepts any linear function with probability of at least $c$.
A test has \emph {perfect completeness} if $c=1$. A linearity test
has \emph{soundness} $s$ if it accepts any function $f$ with
agreement at most $\epsilon$ with all linear functions, with
probability of at most $s + \epsilon'$, where $\epsilon' \to 0$ when
$\epsilon \to 0$. We define the \emph{query complexity} $q$ of a
test as the maximal number of queries it performs. In the case of
the BLR test, it has perfect completeness, soundness $s=1/2$ (with
$\epsilon'=2 \epsilon$) and query complexity $q=3$.

If one repeats a linearity test with query complexity $q$ and
soundness $s$ independently $t$ times, the query complexity grows to
$q'=qt$ while the soundness reduces to $s'=s^t$. So, it makes sense
to define the \emph{amortized query complexity} $\bar{q}$ of a test
as $\bar{q} = q / \log_2{(1/s)}$. Independent repetition of a test
doesn't change it's amortized query complexity. Notice that the BLR
test has amortized query complexity $\bar{q}=3$.

Linearity tests are a key ingredient in the PCP theorem, started in
the works of Arora and Safra  \cite{AS98} and Arora, Lund, Motwani,
Sudan and Szegedy \cite{ALM+98}. In order to improve PCP
constructions, linearity tests were studied in order to improve
their amortized query complexity.

Samorodnitsky and Trevisan \cite{ST00} have generalized the basic
BLR linearity test. They introduced the \emph{Complete Graph Test}.
The Complete Graph Test (on $k$ vertices) is:
\begin{enumerate}
    \item Choose $\x_1,...,\x_k \in \{0,1\}^n$ independently
    \item Verify $f(\x_i + \x_j) = f(\x_i) + f(\x_j)$ for all $i,j$
\end{enumerate}
This test has perfect completeness and query complexity $q={k
\choose 2} + k$. They show that all the ${k \choose 2}$ tests that
the Complete Graph Test performs are essentially independent, i.e.
that the test has soundness $s = 2^{-{k \choose 2}}$. This makes
this test have amortized query complexity $\bar{q} = 1 +
\theta(1/\sqrt{q})$. They show that this test is optimal among the
family of Hyper-Graph Tests (see \cite{ST00} for definition of this
family of linearity tests), and raise the question of whether the
Complete Graph Test is optimal among all linearity tests, i.e. does
a test with the same query complexity but with better soundness
exist?

They partially answer this question in \cite{ST06}, where (among
many other results) they show that no non-adaptive linearity test
can perform better than the Complete Graph Test. A test is called
\emph{non-adaptive} if it first chooses $q$ locations in the truth
table of $f$, then queries them, and based on the results accept or
rejects $f$. Otherwise, a test is called \emph{adaptive}. An
adaptive test may decide on its query locations based on the values
of $f$ in previous queries.

The proof technique of \cite{ST06} uses the algebraic analysis of
the Gowers Norm of certain functions. The Gowers Norm is a measure
of local closeness of a function to a low degree polynomial. For
more details regarding the definition and properties of the Gowers
Norm, see \cite{GT05} and \cite{Sam07}.

Ben-Sasson, Harsha and Raskhodnikova prove in \cite{BHR05} that any
adaptive linearity test with completeness $c$, soundness $s$ and
query complexity $q$ can be transformed into a non-adaptive
linearity test with the same query complexity, perfect completeness
and soundness $s' = s + 1-c$. Combining their result with the result
of \cite{ST06} proves the lower bound also for adaptive linearity
tests.

We also prove the same optimal lower bound for adaptive linearity
test, but our proof technique is arguably simpler and more direct
than the one used in \cite{ST06}. We also study, like \cite{ST06},
the behavior of linearity tests on quadratic functions. However,
instead of employing algebraic analysis of the Gowers Norm of
certain functions, we provide a more direct combinatorial proof,
studying the behavior of linearity tests on random quadratic
functions. This proof technique also lets us prove directly the
lower bound also for adaptive linearity tests.

\subsection{Our techniques}
We model adaptive tests using test trees. A test tree $T$ is a
binary tree, where in each inner vertex $v$ there is some label
$\x(v) \in \{0,1\}^n$, and the leaves are labeled with either
\emph{accept} or \emph{reject}. Running a test tree on a function
$f$ is done by querying at each stage $f$ on the label of the
current vertex (starting at the root), and following one of the two
edges leaving the vertex, depending on the query response. When
reaching a leaf, its label (\emph{accept} or \emph{reject}) is the
value of that $f$ gets in $T$ . An adaptive test $\T$ can always be
modeled as first randomly choosing a test tree from some set
$\{T_i\}$, according to some distribution on the test trees, then
running the test tree on $f$.

It turns out that in order to prove a lower bound which matches the
upper bound of the Complete Graph Test, it is enough to consider
functions $f$ which are quadratic. Actually, it's enough to consider
$f$ which is a random quadratic function.

A function $f$ is quadratic if it can be presented as
$f(x_1,...,x_n) = \displaystyle\sum_{i,j} a_{i,j} x_i x_j +
\displaystyle\sum_i b_i x_i + c$ for some values $a_{i,j},b_i,c \in
\F_2$. We study the behavior of running test trees on a random
linear function, and on a random quadratic function.

The main idea is as follows. Let $v$ be some inner vertex in a test
tree $T$, with the path from the root of $T$ to $v$ being
$v_0,...,v_{k-1},v$. If $\x(v)$ is linearly dependent on
$\x(v_0),...,\x(v_{k-1})$, then when running $T$ on any linear
function, the value of $f(\x(v))$ can be deduced from the already
known values of $f(\x(v_0)),...,f(\x(v_{k-1}))$. Therefore, if the
vertex $v$ is reached, then the same edge leaving $v$ will always be
taken by any linear function. Additionally, if $\x(v)$ is linearly
independent of $\x(v_0),...,\x(v_{k-1})$, then either $v$ is never
reached running $T$ on linear functions, or the two edges leaving
$v$ are taken with equal probability when running $T$ on a random
linear function. A similar analysis can be made when running $T$ on
quadratic functions, replacing \emph{linear dependence} with a
corresponding notion of \emph{quadratic dependence}.

Using this observation, we can define the \emph{linear rank} of a
leaf $v$, marked $l(v)$, as the linear rank of labels on the path
from the root to $v$. We prove that running the test tree $T$ on a
random linear function reaches $v$ with probability $2^{-l(v)}$.
Similarly, we define the \emph{quadratic rank} of a leaf $v$, marked
$q(v)$, as the quadratic rank of those labels, and we proving that
running $T$ on a random quadratic function reaches $v$ with
probability $2^{-q(v)}$. We prove that the quadratic rank of any set
cannot be much larger than its linear rank, and in particular that
$q(v) \le {l(v) \choose 2} + l(v)$ for all leaves $v$. We use this
inequality to prove that a test which has completeness $c$ and query
complexity $q$ accepts a random quadratic function with a
probability of at least $c - 1 + 2^{-q + \phi(q)}$, where $\phi(q)$
is defined as the unique non-negative solution to ${\phi(q) \choose
2} + \phi(q) = q$.

We use this to show that any linearity test with completeness $c$
and query complexity $q$ must have $s \ge 2^{-q + \phi(q)}$. In
particular, the Complete Graph Test on $k$ vertices has perfect
completeness, soundness $s = 2^{-{k \choose 2}}$ and query
complexity $q = {k \choose 2} + k$. Since $\phi(q)=k$ the Complete
Graph Test is optimal among all adaptive tests with the same query
complexity.

In fact, we prove a stronger claim. We say that a test $\T$ has
\emph{average query complexity} $q$ if for any function $f$, the
average number of queries performed is at most $q$. In particular
any test with query complexity $q$ also has average query complexity
$q$. We prove that for any test with completeness $c$ and average
query complexity $q$, the soundness is at least $s \ge 2^{-q +
\phi(q)}$.

We present and analyze linearity tests over $\F_2$. Linearity tests
can also be considered over larger fields or groups. Our lower bound
actually generalizes easily to any finite field, but for ease of
presentation, and since the techniques are exactly the same, we
present everything over $\F_2$. We comment further on the
modifications required for general finite fields in
Section~\ref{prelim}.

\section{Preliminaries}\label{prelim}
\subsection{Linearity tests}
We call a function $f:\{0,1\}^n \to \{0,1\}$ linear if it can be
written as $f(x_1,...,x_n) = a_1 x_1 + ... + a_n x_n$ for some
$a_1,...,a_n \in \{0,1\}$ where addition and multiplication are in
$\F_2$.

A linearity test is a randomized algorithm with oracle access to the
truth table of $f$, which is supposed to distinguish the following
two extreme cases:
\begin{enumerate}
  \item $f$ is linear (accept)
  \item $f$ is $\epsilon$-far from linear functions (reject)
\end{enumerate}
where the agreement of two functions $f,g:\{0,1\} \to \{0,1\}$ is
defined as $d(f,g) = |Pr_{\x}[f(\x)=g(\x)] - Pr_{\x}[f(\x) \ne
g(\x)]|$, and $f$ is $\epsilon$-far from linear functions if the
agreement it has with any linear function is at most $\epsilon$.

We now follow with some standard definition regarding linearity
tests (or more generally, property tests). We say a test has
\emph{completeness} $c$ if for any linear function $f$ the test
accepts with probability at least $c$. A test has \emph{perfect
completeness} if $c=1$. We say a test has \emph{soundness} $s$ if
for any $f$ which is $\epsilon$-far from linear the test accepts
with probability at most $s + \epsilon'$, where $\epsilon' \to 0$
when $\epsilon \to 0$ (in fact, we talk about a family of linearity
tests, for $n \to \infty$, but we ignore this subtle point).

A test is said to have \emph{query complexity} $q$ if it accesses
the truth-table of $f$ at most $q$ times (for any choice of it's
internal randomness). A test is said to have \emph{average query
complexity} $q$ if for any function $f$, the average number of
accesses (over the internal randomness of the test) done to the
truth table of $f$ is at most $q$. Obviously, any test with query
complexity $q$ is also a test with average query complexity $q$.

We say a test is \emph{non-adaptive} if it chooses all the locations
it's going to query in the truth table of $f$ before reading any of
their values. Otherwise, we call the test \emph{adaptive}.

We now turn to model adaptive tests in a way that will be more
convenient for our analysis. We first define a test tree and running
a test tree on a function.

\begin{defi} A \emph{test tree} on functions $\{0,1\}^n \to \{0,1\}$
is a rooted binary tree $T$.
On each inner vertex of the tree $v$ there is a label $x(v) \in
\{0,1\}^n$. On each leaf there is a label of either \emph{accept} or
\emph{reject}.
\end{defi}

\begin{defi} \emph{Running a test tree $T$ on a function $f$} is done
as follows. We start at the root of
the tree $v_0$, read the value of $f(x(v_0))$, and according to the
value take the left or the right edge leaving $v_0$. We continue in
this fashion on inner vertices of $T$ until we reach a leaf of $T$.
The \emph{value of $f$ in $T$} is the value of the end leaf (i.e.
\emph{accept} or \emph{reject}), and the \emph{depth of $f$ in $T$}
is the depth of the end vertex of $f$ in $T$.
\end{defi}

Using these definitions, we can now model adaptive tests. We
identify an adaptive test $\T$ on functions $\{0,1\}^n \to \{0,1\}$
with a distribution of binary trees $\{T_i\}$ (also on functions
$\{0,1\}^n \to \{0,1\}$). Running the test $\T$ on a function $f$ is
done by randomly choosing one of the trees $T_i$ (according to their
distribution), and then running the test tree $T_i$ on $f$. The
result of the function $f$ in the test tree $T_i$ is the result the
test $\T$ returns on $f$.

Notice that a test has query complexity $q$ iff all trees $T_i$ has
depth at most $q$, and has average query complexity $q$ iff for any
function $f$, the average depth reached in a random tree from
$\{T_i\}$ is at most $q$.

In order to define our main theorem, we will define the following
function. For $x>0$ define $\phi(x)$ as the unique real positive
solution to $\phi(x)^2/2 + \phi(x)/2 = x$. Notice that for positive
integer $\phi(x)$, this is the same as ${\phi(x) \choose 2} +
\phi(x) = x$. The following is the main theorem of this paper:

\begin{thm}\label{maintheorem}(main theorem)
Let $\T$ be an adaptive test with completeness $c$, soundness $s$
and average query complexity $q \ge 1$. Then $s + 1- c \ge 2^{-q +
\phi(q)}$.
\end{thm}

Notice that for large $q$, $\phi(q) \approx \sqrt{2q}$, also
$\sqrt{q} \le \phi(q) \le \sqrt{2q}$, so we get that in particular,
$s + 1 -c \ge 2^{-q + \theta(\sqrt{q}})$.

The Complete Graph Test was presented in \cite{ST00}. The test (on a
graph with $k$ vertices) can be described as choosing
$\x_1,...,\x_k$ at random, and querying $f$ at $\x_i$ (for $i=1..k$)
and on $\x_i + \x_j$ (for $1 \le i < j \le k$). The test accepts $f$
if for any $i,j$
$$f(x_i) + f(x_j) + f(x_i+x_j) = 0$$

In \cite{ST00} it is proven that the Complete Graph Test has perfect
completeness and soundness $s=2^{-{k \choose 2}}$. The total number
of queries performed is $q = k + {k \choose 2}$, so by our
definitions, $k = \phi(q)$ and $s = 2^{-q + \phi(q)}$. We have the
following corollary:

\begin{cor}
The Complete Graph Test is optimal among all adaptive linearity
tests.
\end{cor}

\begin{rem}
We state and prove all results for functions $f:\{0,1\}^n \to
\{0,1\}$. In fact, the lower bound result on adaptive linearity
tests holds for functions $f:\F^n \to \F$ for any finite field $\F$,
and not just $\F_2$, with only minor adjustments to the definitions
and proofs. We need to make the following modifications:
\begin{enumerate}
    \item Define "$\epsilon$-far from linear functions" for general
fields
    \item Test trees should have $|F|$ edges leaving each edge
instead of $2$
    \item The proof that random quadratic functions are far from
linear, proved in Section~\ref{quadfarfromlinlemmaproofsection},
should be slightly modified
\end{enumerate}
Since the results follow simply for any finite field, we chose to
present the results over $\F_2$ to make the presentation simpler and
clearer.
\end{rem}

\section{Quadratic functions}\label{quadraticfunctionssection}
We will see that in order to prove Theorem~\ref{maintheorem}, it
will be enough to limit the functions $f$ to be quadratic. We say a
function $f$ is quadratic if it can be written as:
$$ f(x_1,...,x_n) = \displaystyle\sum_{i,j} a_{i,j} x_i x_j +
\displaystyle\sum_i b_i x_i + c$$ for some $a_{i,j},b_i,c \in \F_2$.

In fact, for our usage, we will force our quadratic functions $f$ to
have $f(0)=0$ (equivalently, $c=0$ in the above description). So,
throughout this paper, when speaking of quadratic functions, we
actually speak of quadratic functions $f$ with the added condition
$f(0)=0$.

We will study the dynamics of a test tree $T$ in a linearity test
$\T$, in two cases - when applied to a uniformly random linear
function, and when applied to a uniformly random quadratic function.

The following technical lemma is the key ingredient to the proof of
the Theorem~\ref{maintheorem}.

\begin{lemma}\label{mainlemma}
Let $\T$ be an adaptive linearity test with completeness $c$ and
average query complexity $q$. Then running $\T$ on a random
quadratic function returns \emph{accept} with probability at least
$c - 1 + 2^{-q + \phi(q)}$.
\end{lemma}

In order to prove Theorem~\ref{maintheorem}, we will also need the
following simple lemma:
\begin{lemma}\label{randquadfarfromlinearlemma}
Let $f$ be a random quadratic function. Then the probability that
$f$ is not $2^{-\Omega(n)}$-far from linear functions is
$2^{-\Omega(n)}$.
\end{lemma}

Theorem~\ref{maintheorem} now follows directly from
Lemmas~\ref{mainlemma} and~\ref{randquadfarfromlinearlemma}. We
sketch now it's proof following the two lemmas.

\proof(of the main theorem)
The average probability that $\T$ returns \emph{accept} on a random
quadratic function which is $2^{-\Omega(n)}$-far from linear
functions is at least $c - 1 + 2^{-q + \phi(q)} - 2^{-\Omega(n)}$.
So, there exists some quadratic function $f$ which is
$2^{-\Omega(n)}$-far from linear and on which $\T$ returns
\emph{accept} with probability at least $c - 1 + 2^{-q + \phi(q)} -
2^{-\Omega(n)}$. Taking $n \to \infty$ shows that $s + 1 - c \ge
2^{-1 + \phi(q)}$.
\qed

The remainder of the paper is organized as follows.
Lemma~\ref{mainlemma} is proved in
Section~\ref{mainlemmaproofsection}, and
Lemma~\ref{randquadfarfromlinearlemma} is proved in
Section~\ref{quadfarfromlinlemmaproofsection}.

\section{Linearity test applied to a random quadratic
function}\label{mainlemmaproofsection}
We study tests and test trees applied to linear and quadratic
functions, in order to prove Lemma~\ref{mainlemma}. Let $\T$ be an
adaptive test with completeness $c$ and average query complexity
$q$. Let $T$ be a some test tree which is a part of the test $\T$.

We start by studying the dynamics of applying $T$ to linear
functions. Assume we know that $f$ is a linear function, and we are
at some vertex $v \in T$, where the path from the root to $v$ is
$v_0,..,v_{k-1},v$. Assume $\x(v)$ is linearly dependant on
$\x(v_0),...,\x(v_{k-1})$. Since we know $f$ is linear, we can
deduce the value of $\x(v)$ from $\x(v_0),...,\x(v_{k-1})$, and so
we will always follow the same edge leaving $v$ when we apply $T$ to
any linear function. On the other hand, if $\x(v)$ is linearly
independent of $\x(v_0),...,\x(v_{k-1})$, we know that when we apply
$T$ to a random linear function, either we never reach $v$, or we
have equal chances of taking any of the two edges leaving $v$.

This gives rise to the following formal definition:
\begin{defi}
Let $v$ be a leaf in $T$, where the path from the root to $v$ is
$v_0,v_1,...,v_{k-1},v$. We define the \emph{linear degree} of $v$,
marked $l(v)$, to be the linear rank of $\x(v_0),...,\x(v_{k-1})$.
\end{defi}

We define $L_T$ to be the set of leaves of $T$ to which linear
functions can arrive. i.e, $v \in L$ if the path from the root to
$v$, $v_0,...,v_{k-1},v$ always takes the "correct" edge leaving any
vertex $v_i$ with $\x(v_i)$ linearly dependent on
$\x(v_0),...,\x(v_{i-1})$.

The following lemma formalizes the discussion above:
\begin{lemma}\label{treelinearlemma}
For any test tree $T$:
\begin{enumerate}
  \item For any $v \in L_T$, the probability that a random linear
function will arrive to
  $v$ is $2^{-l(v)}$
  \item $\displaystyle\sum_{v \in L_T} 2^{-l(v)} = 1$
\end{enumerate}
\end{lemma}

For $v \in L_T$, we define $c(v)$ to be $1$ if the value of $v$ is
\emph{accept}, and $c(v)=0$ otherwise. Since the completeness of
$\T$ is $c$, we have that the probability that $\T$ returns
\emph{accept} on a random linear function is at least $c$. On the
other hand, for any test tree $T$ in $\T$, the probability that a
random linear function will return \emph{accept} is exactly
$\displaystyle\sum_{v \in L_T} c(v) 2^{-l(v)}$. So, the following
lemma follows:

\begin{lemma}\label{avgtreelinearlemma}
$\E_T \displaystyle\sum_{v \in L_T} c(v) 2^{-l(v)} \ge c$
\end{lemma}
where by $\E_T$ here and throughout the paper we mean the average
value of a random test tree $T$ in $\T$.

We now generalize the concept of linear dependence to quadratic
functions.

\begin{defi}
Let $\x_1,...,\x_k \in \{0,1\}^n$.
\begin{enumerate}
  \item We say $\x_1,...,x_k$ are \emph{quadratically dependent} if
there are constants $a_1,...,a_k \in \F_2$, not all zero, s.t. for
any quadratic function $f$ we have: $a_1 f(\x_1) + ... + a_k f(\x_k)
= 0$. otherwise will call $x_1,...,x_k$ \emph{quadratically
independent}.

  \item We say $\x_k$ is \emph{quadratically dependent} on
$\x_1,...,\x_{k-1}$ if there are constants $a_1,...,a_{k-1} \in
\F_2$ s.t. for any quadratic function $f$ we have: $f(\x_k) = a_1
f(\x_1) + ... + a_{k-1} f(\x_{k-1})$. Otherwise we say $\x_k$ is
\emph{quadratically independent} of $\x_1,...,\x_{k-1}$.

  \item We define the \emph{quadratic dimension} of
$\x_1,...,\x_k$ to be the size of the largest subset of
$\{\x_1,...,\x_k\}$ which is quadratically independent.
\end{enumerate}
\end{defi}

This definition may seem obfuscated, but the following alternative
yet equivalent definition will clarify it. The space of quadratic
functions over $\{0,1\}^n$ is a linear space over $\F_2$. Let $M$ be
it's generating matrix, i.e. the rows of $M$ are a base for the
linear space (in particular, the dimensions of $M$ are $({n \choose
2} + n) \times 2^n$). A column of $M$ corresponds to an input $\x
\in \{0,1\}^n$. Now, $\x_1,...,\x_k$ are quadratically dependent iff
the columns corresponding to them are linearly dependent, and
similarly for the other definitions.

Notice that the usual definition of linear dependence is equivalent
to this more complex definition, when applied to the linear space of
all linear functions.

We now can repeat the informal discussion at the start of this
section, except this time for quadratic functions, with all the
reasoning left intact. Let $v \in T$ be a vertex, with path from the
root being $v_0,...,v_{k-1},v$. Assume $\x(v)$ is quadratically
dependent on $\x(v_0),...,\x(v_{k-1})$, and $f$ is any quadratic
function. The value of $f(\x(v))$ can be deduced from the already
known values of $f(\x(v_0)),...,f(\x(v_{k-1}))$, and so only one
edge leaving $v$ will be taken on all quadratic functions.
Alternatively, if $x(v)$ is quadratically independent on
$\x(v_0),...,\x(v_{k-1})$, then a random quadratic function either
never reaches $v$, or has equal chances of taking each of the two
edges leaving $v$.

This leads to the following definition and lemma for quadratic
degree of a vertex $v \in T$, similar to the ones for linear degree.

\begin{defi}
Let $v$ be a leaf in $T$, where the path from the root to $v$ is
$v_0,v_1,...,v_{k-1},v$. We define the \emph{quadratic degree} of
$v$, marked $q(v)$, to be the quadratic rank of
$\x(v_0),...,\x(v_{k-1})$.
\end{defi}

We define $Q_T$ to be the set of leaves of $T$ to which quadratic
functions can arrive. Naturally $L_T \subseteq Q_T$. The following
lemma on quadratic degree follows from the discussion above:

\begin{lemma}\label{lemmalintree}
For any test tree $T$:
\begin{enumerate}
  \item For any $v \in Q_T$, the probability that a random quadratic
function will arrive to
  $v$ is $2^{-q(v)}$
  \item $\displaystyle\sum_{v \in Q} 2^{-q(v)} = 1$
  \item For any $v \in L_T$ we have $q(v) \ge l(v)$
\end{enumerate}
\end{lemma}

Last, we mark the depth of a vertex $v \in T$ by $d(v)$. Since $\T$
has average query complexity $q$, we know that for any function $f$,
the average depth of running a random tree $T$ of $\T$ on $f$ is at
most $q$. So, this also holds for a random linear function. However,
the average depth a random linear function arrives on a tree $T$ is
exactly $\displaystyle\sum d(v) 2^{-l(v)}$, so the following lemma
follows.

\begin{lemma}\label{avgdepthlinlemma}
$\E_T \displaystyle\sum_{v \in L_T} d(v) 2^{-l(v)} \le q$
\end{lemma}

We now wish to make a connection between $q(v)$ and $l(v)$ for
vertices $v \in L_T$.

First, we prove that following lemma:
\begin{lemma} \label{quadsumlemma}
For any $\x_1,...,\x_k \in \{0,1\}^n$ there are coefficients
$a_{i,j}, b_i \in \F_2$ s.t. for any quadratic function $f$ we have:
$$f(\x_1 + ... + \x_k) = \displaystyle\sum_{i,j} a_{i,j} f(\x_i +
\x_j) +
\displaystyle\sum_i b_i f(\x_i)$$
\end{lemma}

\proof
Let $f(\x)$ by some polynomial of degree $d$. It's derivative in the
$\y$ direction is defined to be $f_{\y}(\x) = f(\x+\y) - f(\x)$.
It's easy to see that the degree of $f_{\y}$ as a function of $\x$
is at most $d-1$. So, taking 3 derivatives from a quadratic function
makes it the zero function, and so in particular for any quadratic
function f, we we take it's derivatives in directions $\x,\y$ and
$\z$, and evaluate the result at $0$, we get that
$$(((f_{\x})_{\y})_{\z} (0) = 0$$

Opening this expression yields:
$$f(\x+\y+\z) - f(\x+\y) - f(\x+\z) - f(\y+\z) + f(\x) + f(\y) +
f(\z) - f(0) = 0$$

Since $f(0)=0$, we can express $f(\x+\y+\z)$ as a sum of application
of $f$ on an element, or sum of two elements in $\{x,\y,\z\}$. This
proves the lemma for $k=3$. For $k>3$ we use simple induction.
\qed

Now we can bound $l(v)$ in term of $q(v)$. We first prove a result
bounding in general the linear rank of a set by it's quadratic rank.

\begin{lemma}\label{connectquadlingenlemma}
Let $\{\x_1,...,\x_k\}$ be elements in $\{0,1\}^n$. Let $l$ be the
their linear rank, and $q$ their quadratic rank. Then
$$q \le {l \choose 2}+l$$
\end{lemma}

\proof
Let $S \subset \{\x_1,...,\x_k\}$ be a maximal quadratic independent
set. $|S| = q$. The linear rank of $S$ is also $l$. Let $S' \subset
S$ be a maximal set of linearly independent elements of $S$.
$|S'|=l$. Assume w.l.o.g that $S' = \{\x_1,...,\x_l\}$. Since every
$x \in S$ is linearly dependent on $S'$, it can be written as a sum
of some of the elements of $S'$.  Using Lemma~\ref{quadsumlemma}, we
get that for any $\x \in S$ there exists coefficients
$a^{(\x)}_{i,j},b^{(\x)}_i \in \F_2$ s.t for any quadratic function
$f$:
$$f(\x) = \displaystyle\sum_{1 \le i < j \le l} a^{(\x)}_{i,j} f(\x_i
+ \x_j) +
\displaystyle\sum_{1 \le i \le l} b^{(\x)}_i f(\x_i)$$

We have assumed that all the elements of $S$ are quadratically
independent. For this to hold, the above equations in the symbolic
variables $f(\x_i + \x_j)$ and $f(\x_i)$ must be linearly
independent. So the number of equations $q$ must be at most the
number of variables, which is ${l \choose 2} + l$. So, we get that:
$$q = |S| \le {l \choose 2} + l$$
\qed

\begin{lemma}\label{connectquadlinlemma}
For any leaf $v \in L_T$, $l(v) \ge \phi(q(v))$
\end{lemma}

\proof
Let $v_0,...,v_{k-1},v$ be the path in $T$ from the root to $v$. Let
$\x_i = \x(v_i)$ for $i=0..k-1$. Apply lemma
\ref{connectquadlingenlemma} on $\{\x_0,...,\x_{k-1}\}$ to get that
$q(v) \le {l(v) \choose 2} + l(v)$. Reversing this formula, since
$\phi(x)$ is monotone, we get that $l(v) \ge \phi(q(v))$.
\qed

We can now prove our main technical lemma (Lemma~\ref{mainlemma}).
We start with some technical lemmas. We define $\psi(x)$ to be $x -
\phi(x)$ for $x \ge 1$, and $0$ for $x < 1$. Notice that $\psi$ is
continuous, and $\psi(x)=x-\phi(x)$ for any non-negative integer
$x$. Hence, using Lemma~\ref{connectquadlinlemma} we get that:

\begin{lemma}\label{psiboundsqllemma}
For any vertex $v$ in a tree $T$, $q(v) - l(v) \le \psi(q(v))$.
\end{lemma}

\begin{lemma}\label{psimonotoneconvexlemma}
$\psi$ is increasing and convex.
\end{lemma}
\proof Since $\psi$ is continuous and constant for $x \le 1$, it's
enough to prove the
claim for $x > 1$ (for increasing it's clear, and once we've proved
$\psi$ is increasing, it shows it's enough to prove convexity for
$x>1$). We first show $\psi$ is increasing.

For $x>1$, define $y = \phi(x)$, so $x = y^2/2 + y/2$ and $\psi(y) =
y^2/2 - y/2$.
$$\frac{d \psi}{d x} = \frac{d \psi}{d y} \frac {d y}{d x} =
\frac{\frac{d \psi} {d y}} {\frac {d x} {d y}} =
\frac{y-1/2}{y+1/2}$$

If $x>1$ then $y=\phi(x)>1$, hence $\frac{d \psi}{d x} > 0$ for
$x>1$, and so $\psi$ is increasing.

To show that $\psi$ in convex,
$$\frac{d^2 \psi}{d x^2} = \frac{d
\left(\frac{y-1/2}{y+1/2}\right)}{dy} \frac{dy}{dx} =
\frac{1}{(y+1/2)^3} > 0$$
\qed

We are now finally ready to prove Lemma~\ref{mainlemma}.

\proof(of Lemma~\ref{mainlemma})
We need to prove that any test $\T$ with completeness $c$ and
average query complexity $q \ge 1$ accepts a random quadratic
function with probability at least $c-1 + 2^{-\psi(q)}$. Let us mark
the probability the test accepts a random quadratic function by $p$.
Let $p_T$ mark the probability that a tree $T$ accepts a random
quadratic function. $p_T$ is at least the probability that a random
quadratic function reaches a leaf in $L_T$ which is labeled
\emph{accept}. So:
$$p_T \ge \displaystyle\sum_{v \in L_T} c(v) 2^{-q(v)}$$

We now follow to analyze $p = \E_T [p_T]$.
$$ p \ge \E_T [\displaystyle\sum_{v \in L_T} c(v) 2^{-q(v)}] =
\E_T [\displaystyle\sum_{v \in L_T} 2^{-l(v)} c(v)
2^{-q(v)+l(v)}]$$.

We divide the sum in the right side into two parts, $p_0 - p_1$,
with $p_0,p_1 \ge 0$, where:
$$p_0 = \E_T [\displaystyle\sum_{v \in L_T} 2^{-l(v)}
2^{-q(v)+l(v)}]$$.
and
$$p_1 = \E_T [\displaystyle\sum_{v \in L_T} 2^{-l(v)} (1 - c(v))
2^{-q(v)+l(v)}]$$.

We start by analyzing $p_1$. Since for any $v$ always $q(v) \ge
l(v)$ we have:
$$p_1 \le \E_T [\displaystyle\sum_{v \in L_T} 2^{-l(v)} (1 - c(v))]$$

Recall that by Lemma~\ref{lemmalintree} for any tree $T$ we have
$$\displaystyle\sum_{v \in L_T} 2^{-l(v)}=1$$
and by Lemma~\ref{avgtreelinearlemma} we have
$$\E_T [\displaystyle\sum_{v \in L_T} 2^{-l(v)} c(v)] \ge c$$
so we conclude that:
$$p_1 \le 1-c$$

We move to analyze $p_0$. Since $\E_T[\displaystyle\sum_{v \in L_T}
2^{-l(v)}]=1$ and since the function $X \to 2^X$ is concave, we have
by Jensen's inequality that:
$$p_0 \ge 2^{\E_T [\displaystyle\sum_{v \in L_T} 2^{-l(v)}
(-q(v)+l(v))]}$$

Now, we have that $q(v)-l(v) \le \psi(q(v))$ by
Lemma~\ref{psimonotoneconvexlemma}, and also by the same lemma,
since $q(v) \le d(v)$, we get $\psi(q(v)) \le \psi(d(v))$. So we
get:
$$\E_T [\displaystyle\sum_{v \in L_T} 2^{-l(v)} (q(v)-l(v))] \le
\E_T [\displaystyle\sum_{v \in L_T} 2^{-l(v)} \psi(d(v))]$$.

Since by Lemma~\ref{psimonotoneconvexlemma} $\psi$ is convex, we get
that again by Jensen's inequality we get that this is at most
$\psi(\E_T [\displaystyle\sum_{v \in L_T} 2^{-l(v)} d(v)])$. By
Lemma~\ref{avgdepthlinlemma}
$$\E_T [\displaystyle\sum_{v \in L_T} 2^{-l(v)} d(v)] \le q$$
where $q$ is the average query complexity of $\T$. So, we conclude
that $p_0 \ge 2^{-\psi(q)}$, and in total
$$p \ge p_0 - p_1 \ge 2^{-\psi(q)} + c - 1$$
\qed

\section{Random quadratic function is far from
linear}\label{quadfarfromlinlemmaproofsection}
In this section we prove Lemma~\ref{randquadfarfromlinearlemma},
i.e. that a random quadratic function is far from linear. We will
use commonly known facts about quadratic functions.

Any quadratic function can be written as:
$$f(\x) = \x^t A \x + <\x,b>$$

The correlation of $f$ with some linear function $g$ is the $g$-th
Fourier coefficient of $f$. The Fourier coefficients of quadratic
functions are well studied. In particular, it is known that all the
Fourier coefficients of $f$ have the same absolute value, and that
the number of non-zero Fourier coefficients is $2^{rank(A+A^t)}$.
So, in order to show that $f$ has no large correlation with some
linear function, it's enough to show that $B = A + A^t$ has high
rank. In particular, in order to show that $f$ is
$2^{-\Omega(n)}$-far from linear functions, we need to show that $B$
has rank $\Omega(n)$. We will show that the probability that a
random quadratic function has rank less than $n/4$ is
$2^{-\Omega(n)}$. We will use the following lemma:

\begin{lemma}\label{matricesnumlemma}
The number of matrices of rank at most $k$ is at most $n^k 2^{nk}$.
\end{lemma}

Using Lemma~\ref{matricesnumlemma}, it's easy to prove
Lemma~\ref{randquadfarfromlinearlemma}. The number of matrices of
rank at most $n/4$ is at most $2^{n^2/4 (1 + o(1))}$. For a random
quadratic function, $B$ is a random symmetric matrix with zero
diagonal, and so the probability that $B$ has rank less than $n/4$
is $2^{-n^2/4 (1 + o(1))} = 2^{-\Omega(n)}$.

Now we finish by proving Lemma~\ref{matricesnumlemma}.
\proof
Let $B$ be a matrix of rank at most $k$. There are ${n \choose k}$
options to choose $k$ rows which span the row span of the matrix,
each other row have at most $2^k$ options since it must be in the
row span of $k$ specific rows. So, the number of possibilities for
rank $k$ matrices is at most:
$${n \choose k} {(2^k)}^{n-k} \le n^k 2^{nk}$$
\qed

\section*{Acknowledgement}
I thank my supervisor, Omer Reingold, for
useful comments and for his constant support and interest in the
work. I thank Alex Samorodnitsky and Prahladh Harsha for helpful
discussions.

\end{document}